\documentclass[a4paper,twoside]{article}

\usepackage[hyphens]{url}  
\usepackage{graphicx} 
\usepackage{natbib}  
\usepackage{caption} 
\usepackage{newfloat}
\usepackage{listings}
\DeclareCaptionStyle{ruled}{labelfont=normalfont,labelsep=colon,strut=off} 
\usepackage{booktabs}

\usepackage{multirow}

\usepackage{multicol, latexsym, amsmath, amssymb}
\usepackage{blindtext}
\usepackage{subcaption}
\usepackage{caption}
\usepackage{svg}
\usepackage{amsmath}
\usepackage{graphicx}
\usepackage{hhline,theorem,verbatim,euscript,url}
\usepackage[babel=true]{csquotes}
\usepackage{listings}
\usepackage{amssymb}
\usepackage{amstext}
\usepackage{mathtools}
\usepackage{textcomp}
\usepackage[ruled,vlined,linesnumbered]{algorithm2e}
\SetAlFnt{\footnotesize\sf}

\usepackage{SCITEPRESS}

\begin{document}
\title{Specializing LLM Agents for Natural Language GUI Test Execution 
}

\author{
	\authorname{Ange Bardet \sup{2}, Moncef Zermache \sup{1}, Alexis Thomas \sup{2}, Arthur Gonzalez \sup{2}, Damien Nortier \sup{2}, Tom Millerat \sup{2}, S\'ebastien Salva \sup{1}, Redha Taguelmimt \sup{1}}
	\affiliation{\sup{1}LIMOS - UMR CNRS 6158, Clermont Auvergne University, UCA, Aubière, France}
	\affiliation{\sup{2}Department of Computing, IUT Clermont Auvergne, UCA, Aubière, France }
	\email{ange.bardet@gmail.com, moncef.zermache@doctorant.uca.fr, alexis.thomas@etu.uca.fr, arthur2gonzalez@gmail.com, 
		damien.nortier@etu.uca.fr, tom.millerat@etu.uca.fr, sebastien.salva@uca.fr, redha.taguelmimt@uca.fr
		}
}

\abstract{
Executing natural language (NL) test cases directly on graphical user interfaces (GUIs) with large language model (LLM) agents can reduce the cost of developing and maintaining 
test scripts. However, existing approaches use relatively general-purpose agent architectures, limiting their ability to exploit differences between LLMs across testing capabilities. We address this limitation with a five-agent framework that decomposes GUI test execution into specialized agents 
and enables capability-specific LLM assignment based on measured performance for each task. To obtain reliable capability measurements, we introduce verified interaction with context replacement, which detects silent interaction failures and prevents corrupted GUI states from propagating to subsequent evaluations. Experiments with 11 locally deployable LLMs reveal substantial performance differences across capabilities, which supports the need for optimizing the agent assignement.
}
\keywords{GUI Application, Natural Language Test Case, Functional Testing, Soundness, Execution Consistency, LLM}

\onecolumn \maketitle \normalsize \vfill


\section{Introduction and Preliminaries}

GUI acceptance testing verifies whether applications satisfy expected user requirements. Conventional approaches implement these tests as executable scripts tied to implementation-specific GUI details, such as element locators, making them costly to develop and maintain. Natural language (NL) test cases provide an easier and a more user-oriented alternative, but reliably executing them on GUIs with large language models (LLMs) 
remains challenging.

Recent work~\cite{DBLP:conf/enase/SalvaT26} addresses these challenges through a three-agent framework that separates NL test execution into three capabilities: navigation, readiness checking (whether the GUI is ready to perform the action), and assertion evaluation. Although this separation improves control over the execution process, it treats 
these capabilities 
relatively broadly and uses the same LLM across the different agents. In particular, navigation involves different types of GUI interactions, such as
\emph{click} and \emph{fill}, for which both execution and readiness requirements can differ substantially. 
Moreover, LLMs may exhibit different capabilities across these tasks, making a single-model configuration potentially suboptimal for a heterogeneous testing pipeline.

In this paper, we develop and present 
a 
five-agent architecture, with dedicated agents for each task. 
This decomposition enables the performance of LLMs to be measured separately for each capability and, 
consequently, allows different LLMs to be assigned to different agents according to their measured performance. 
Such capability-specific assignment requires reliable evaluation: a silent navigation error can corrupt the GUI context and propagate to subsequent agent decisions, making measured performance dependent not only on the evaluated capability but also on upstream execution failures.

To address this issue, we introduce \emph{verified navigation with context replacement}. Each navigation step is 
checked against a trusted execution, and when an error is detected, the corrupted GUI context is replaced before subsequent evaluation continues. This prevents navigation failures from contaminating measurements of downstream capabilities. 

\section{Capability-Specialized Multi-Agent Framework}

We introduce a five-agent architecture that decomposes navigation and readiness according to the type of the GUI action. The resulting framework consists of five specialized agents: 

\begin{itemize}
\item \textbf{Navigation agents} ($agent_{cnav}$ and $agent_{fnav}$) handle \emph{click} and \emph{fill} actions, respectively. Given an action $a_i$ and the current GUI $g_{i-1}$, they produce the GUI operations required to execute the action and return the resulting state $g_i$. 
In this work, we consider only these two navigation action types. The architecture can be extended to additional action types by introducing corresponding specialized agents. 
\item \textbf{Readiness agents} ($agent_{cread}$ and $agent_{fread}$) determine whether the GUI is ready for a \emph{click} or \emph{fill} action, respectively, using an LLM.  
\item \textbf{Assertion agent} ($agent_{assert}$) evaluates an assertion $A_j$ on the resulting GUI state using an LLM. 
\end{itemize}

This decomposition enables LLM performance to be measured independently for each capability. 
For each LLM $m$ and capability $c$, 
we estimate the performance and consistency from repeated executions, producing a capability profile that can guide the configuration of the multi-agent system. Rather than requiring a single LLM to perform all testing tasks, our framework allows potentially different LLMs to be selected for the five capabilities based on their measured performance. 
The assignment can also take into account 
practical constraints such as model size, inference cost, and available computational resources. This enables the framework to support heterogeneous LLM configurations adapted to the requirements and resources of a given testing environment.

\section{Verified Navigation and Reliable Capability Evaluation}

Capability-specific evaluation is only meaningful if failures in one capability do not contaminate the evaluation of subsequent capabilities. This is particularly important for navigation. When executing an NL action, a navigation agent may perform the correct operation, explicitly report a failure, or silently reach an incorrect GUI state while reporting success. 
%
In the last case, subsequent agents operate on an incorrect context, and their decisions may be incorrectly classified. Hence, a navigation failure can propagate through the remainder of a test case and falsify the measured performance of readiness and assertion capabilities.

\begin{figure}[t]
	\centering
	\includegraphics[width=0.5\textwidth]{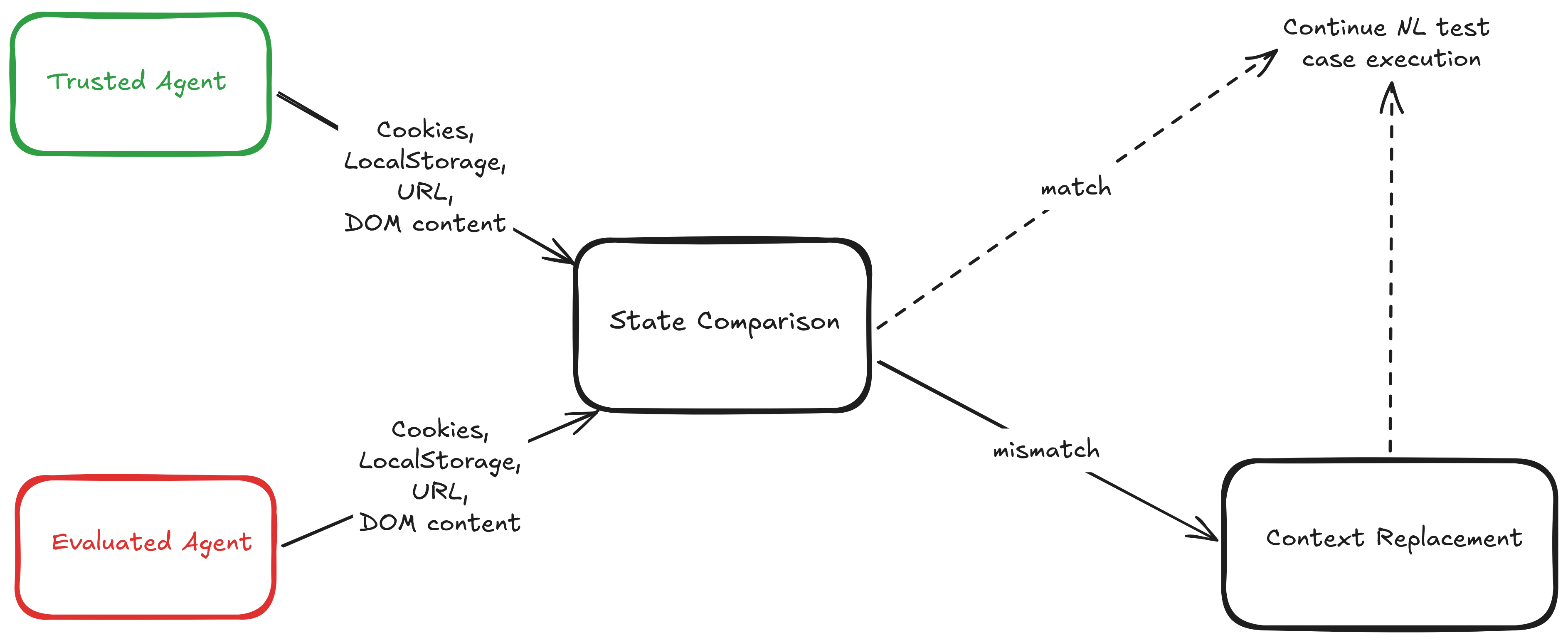}
	\caption{Illustration of a verified navigation with context replacement for reliable capability evaluation.}
	\label{fig:navigation}
\end{figure}

We address this problem with \emph{verified navigation with context replacement}. 
For each navigation action, the evaluated navigation agent and a trusted reference agent (an LLM-as-a-judge) independently execute the same action from the same initial GUI state, producing states $g_e$ and $g_t$, respectively. We compare these states using a GUI equivalence predicate $\mathcal{E}(g_e,g_t)$ based on state information (see Figure~\ref{fig:navigation}). 
If the states are equivalent (i.e., $\mathcal{E}(g_e,g_t)$ holds), the execution continues using the evaluated state. Otherwise, the navigation step is considered faulty and the evaluated agent context is replaced by the trusted agent state: 
\[
g_{\mathrm{next}} =
\begin{cases}
g_e & \text{if } \mathcal{E}(g_e, g_t),\\
g_t & \text{otherwise.}
\end{cases}
\]
Rather than terminating after detecting an incorrect navigation step, context replacement allows the remaining test actions to be evaluated under a valid GUI context. This is important for our goal: we seek to isolate downstream capability measurements from upstream navigation failures and allow the performance of each agent to be evaluated independently.


\section{Experimental Evaluation}

We evaluate the proposed framework using 11 locally deployable LLMs ranging from 3B to 27B parameters, all supporting tool calling, which is required for GUI interaction. LLM inference is performed locally using Ollama on a server equipped with an NVIDIA L40 GPU with 48~GB of memory. 
We conduct 2 experiments to evaluate the capability 
performances and the impact of the proposed navigation evaluation.

\begin{figure}[t]
	\centering
	\begin{subfigure}[b]{0.32\columnwidth}
		\centering
		\includegraphics[width=\columnwidth]{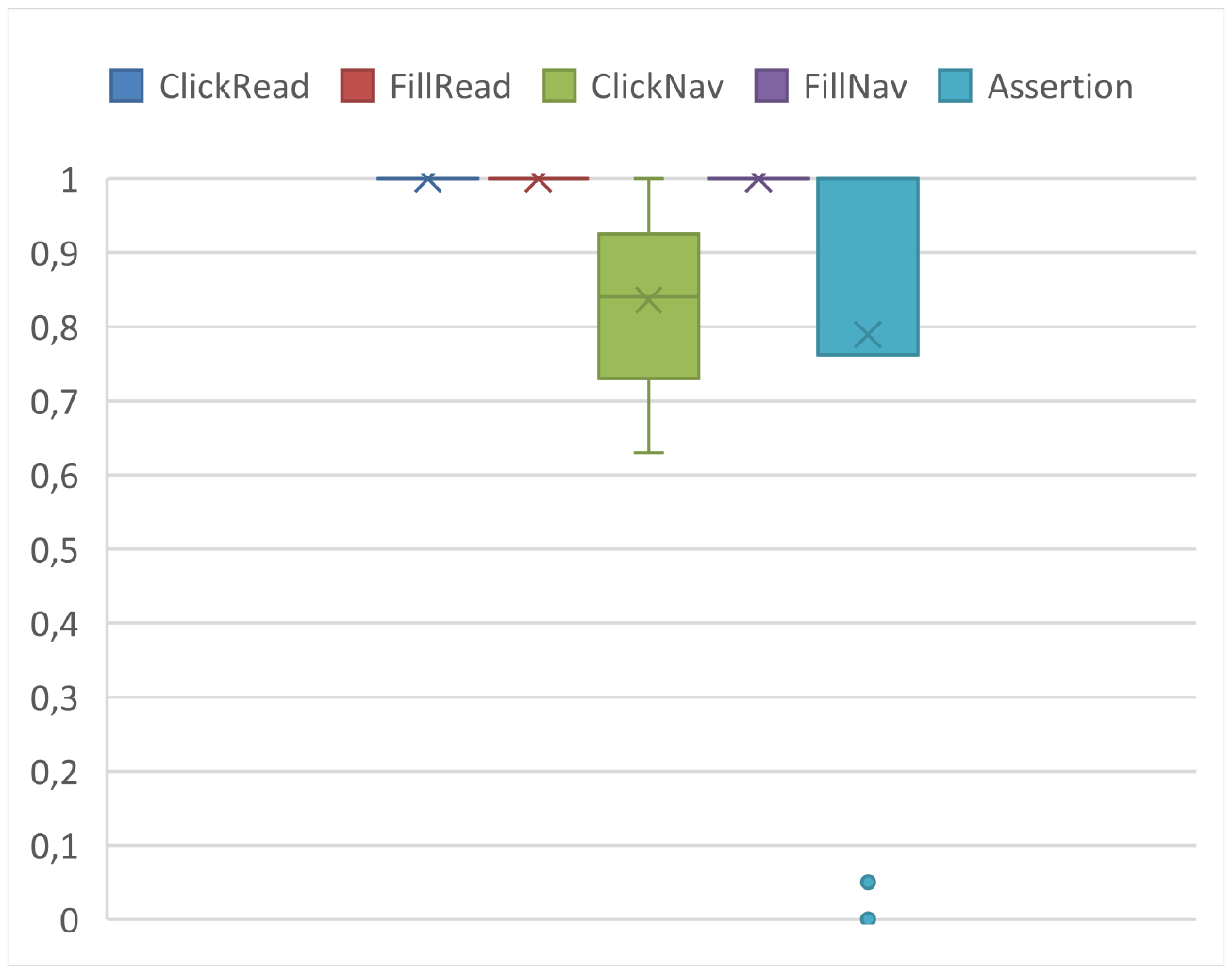}
		\caption{Mistral-nemo 12B}
		\label{fig:sub1}
	\end{subfigure}
	\hfill
	\begin{subfigure}[b]{0.3\columnwidth}
		\centering
		\includegraphics[width=\columnwidth]{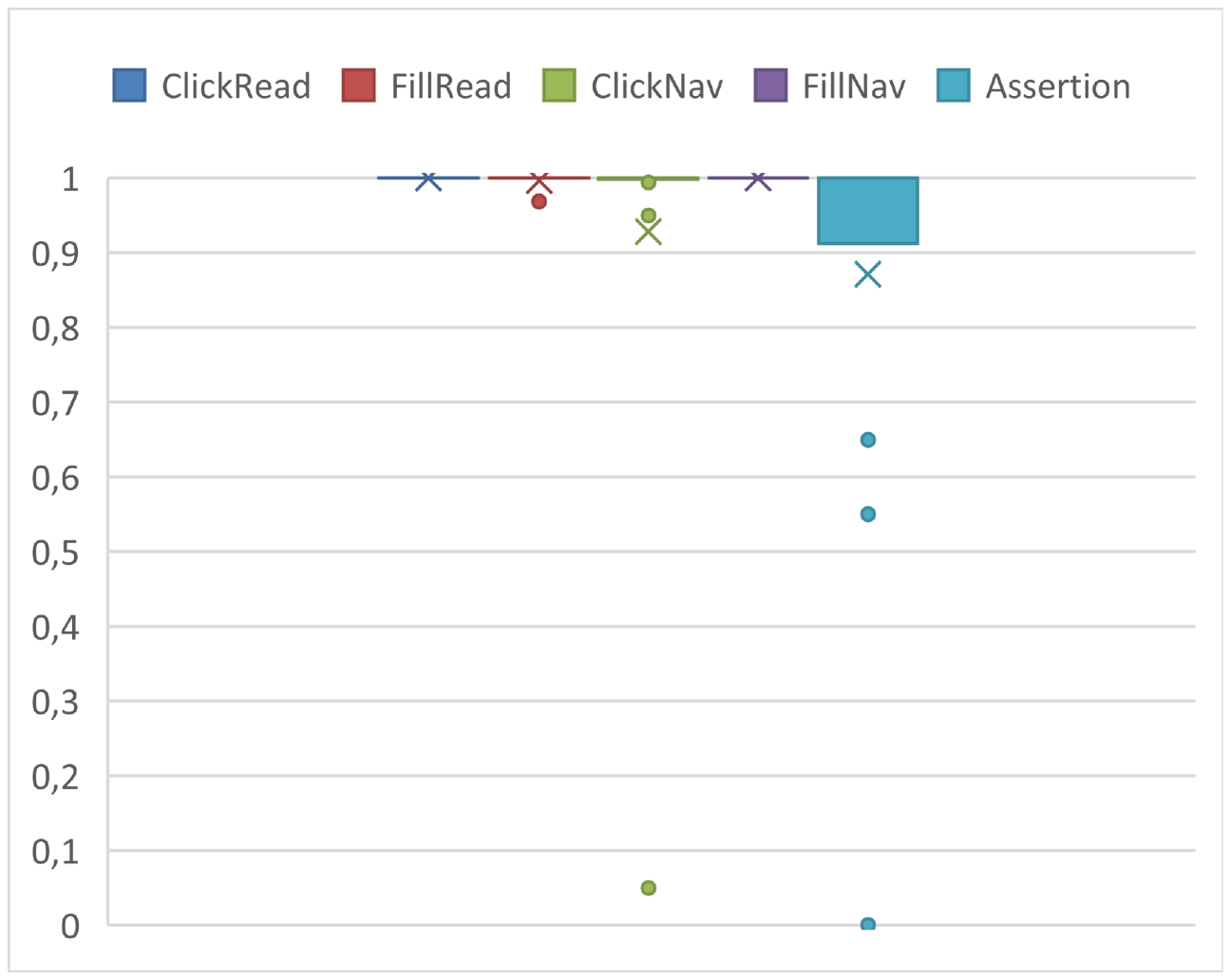}
		\caption{Granite 3.1 8B}
		\label{fig:sub2}
	\end{subfigure}
	\hfill
	\begin{subfigure}[b]{0.3\columnwidth}
		\centering
		\includegraphics[width=\columnwidth]{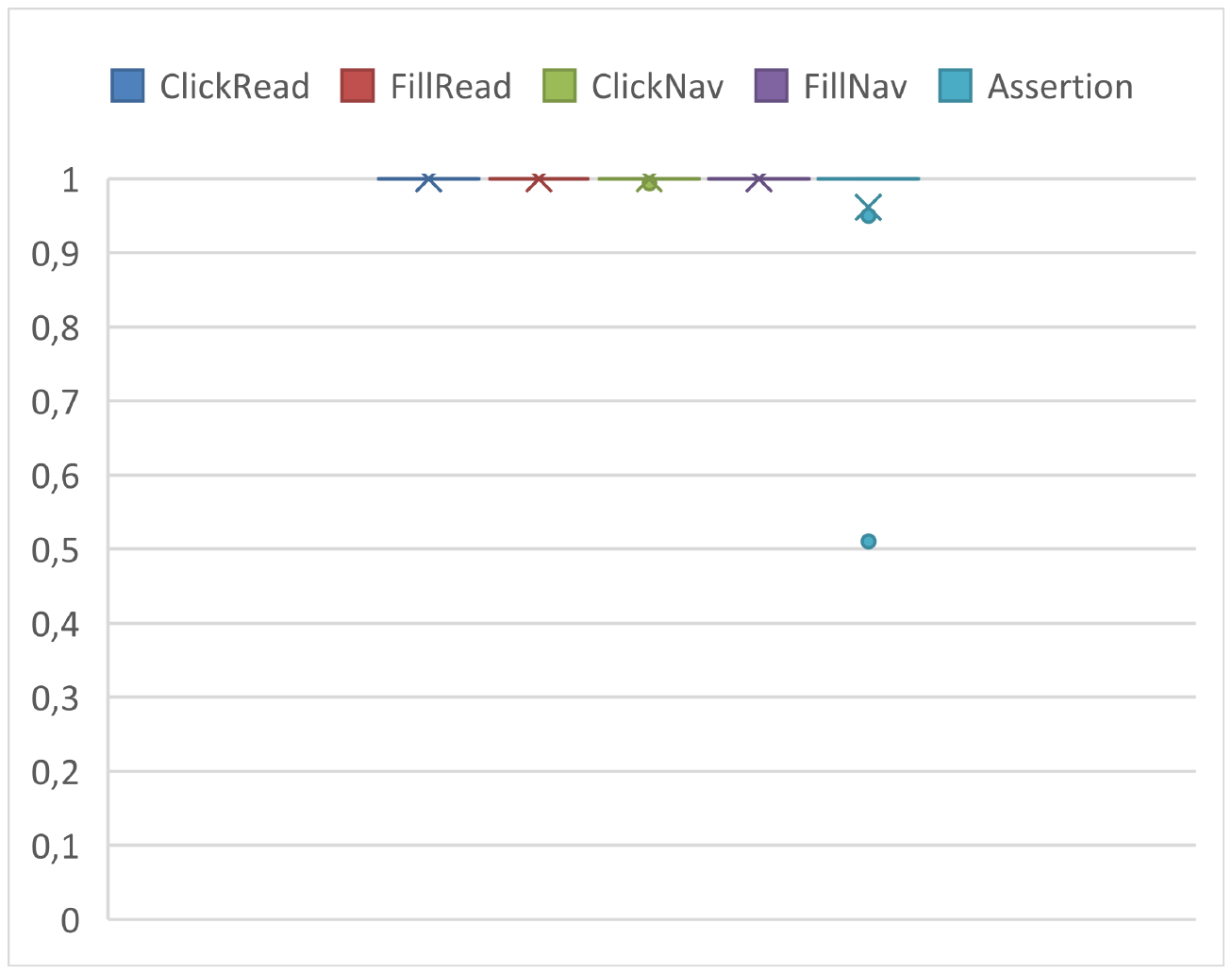}
		\caption{Qwen 3.6 27B}
		\label{fig:sub3}
	\end{subfigure}
	\hfill
	\begin{subfigure}[b]{0.25\columnwidth}
		\centering
		\includegraphics[width=\columnwidth]{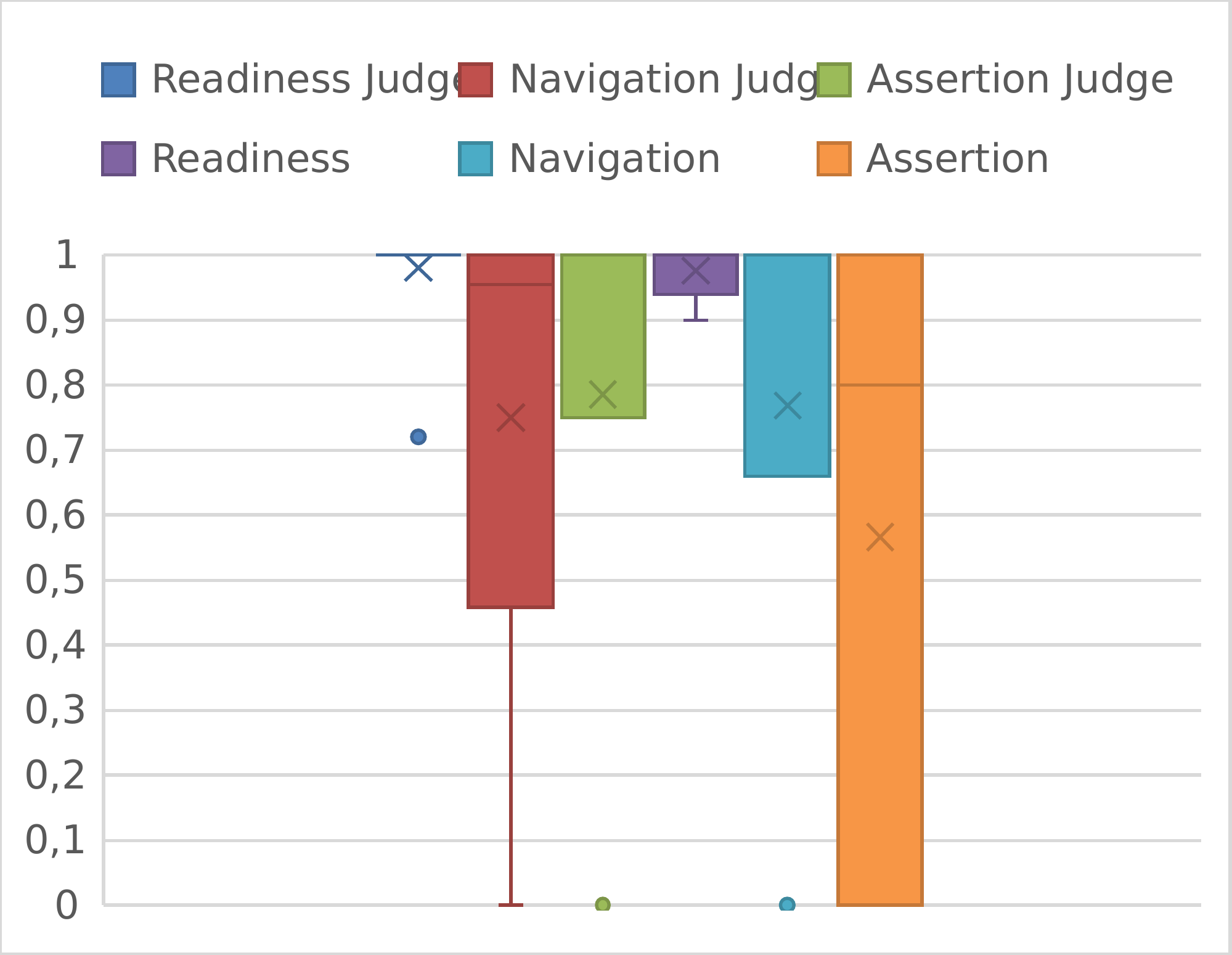}
		\caption{Mistral 7B}
		\label{fig:sub4}
	\end{subfigure}
	\hfill
	\begin{subfigure}[b]{0.25\columnwidth}
		\centering
		\includegraphics[width=\columnwidth]{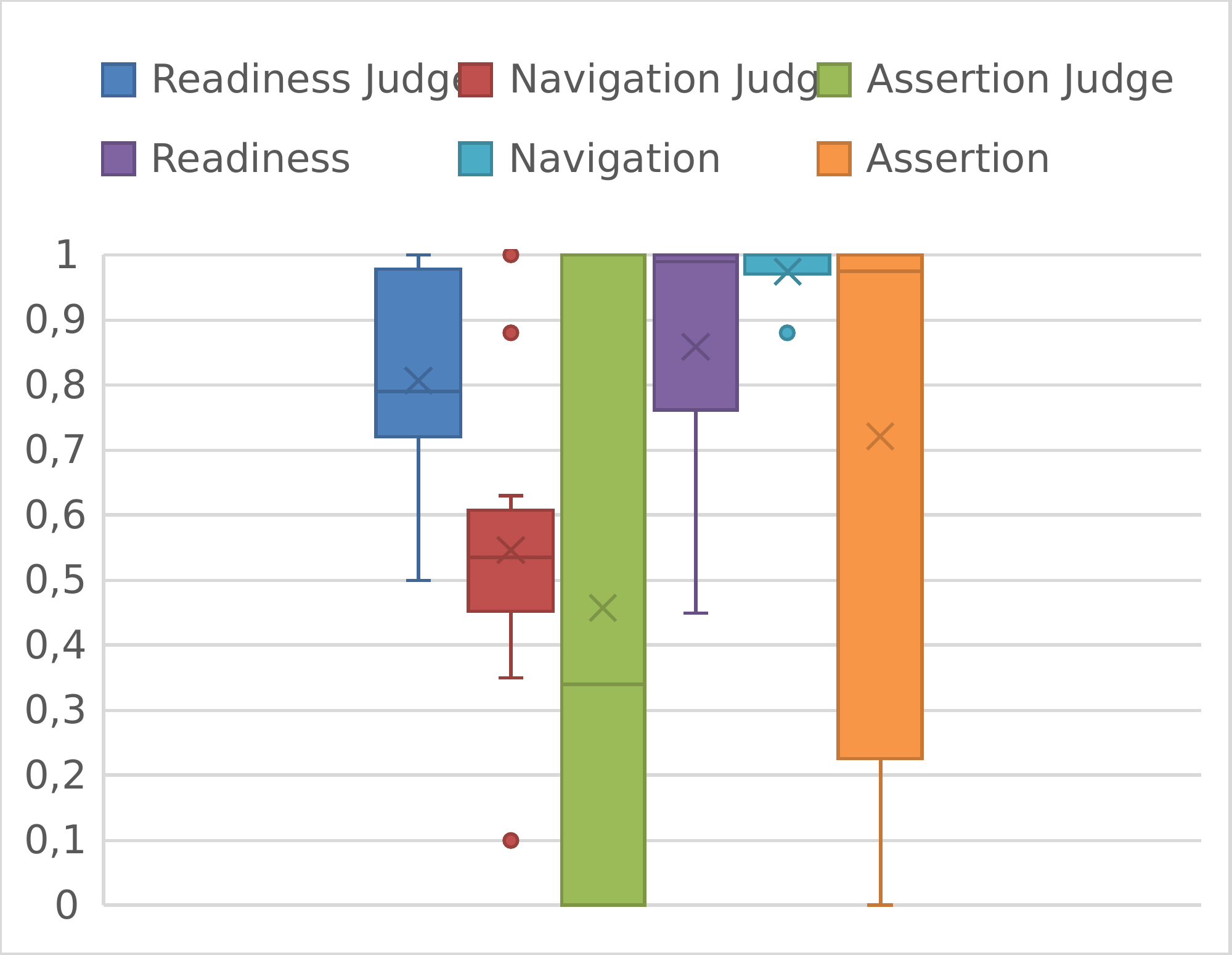}
		\caption{Qwen 2.5 3B}
		\label{fig:sub5}
	\end{subfigure}
	\hfill
	\begin{subfigure}[b]{0.25\columnwidth}
		\centering
		\includegraphics[width=\columnwidth]{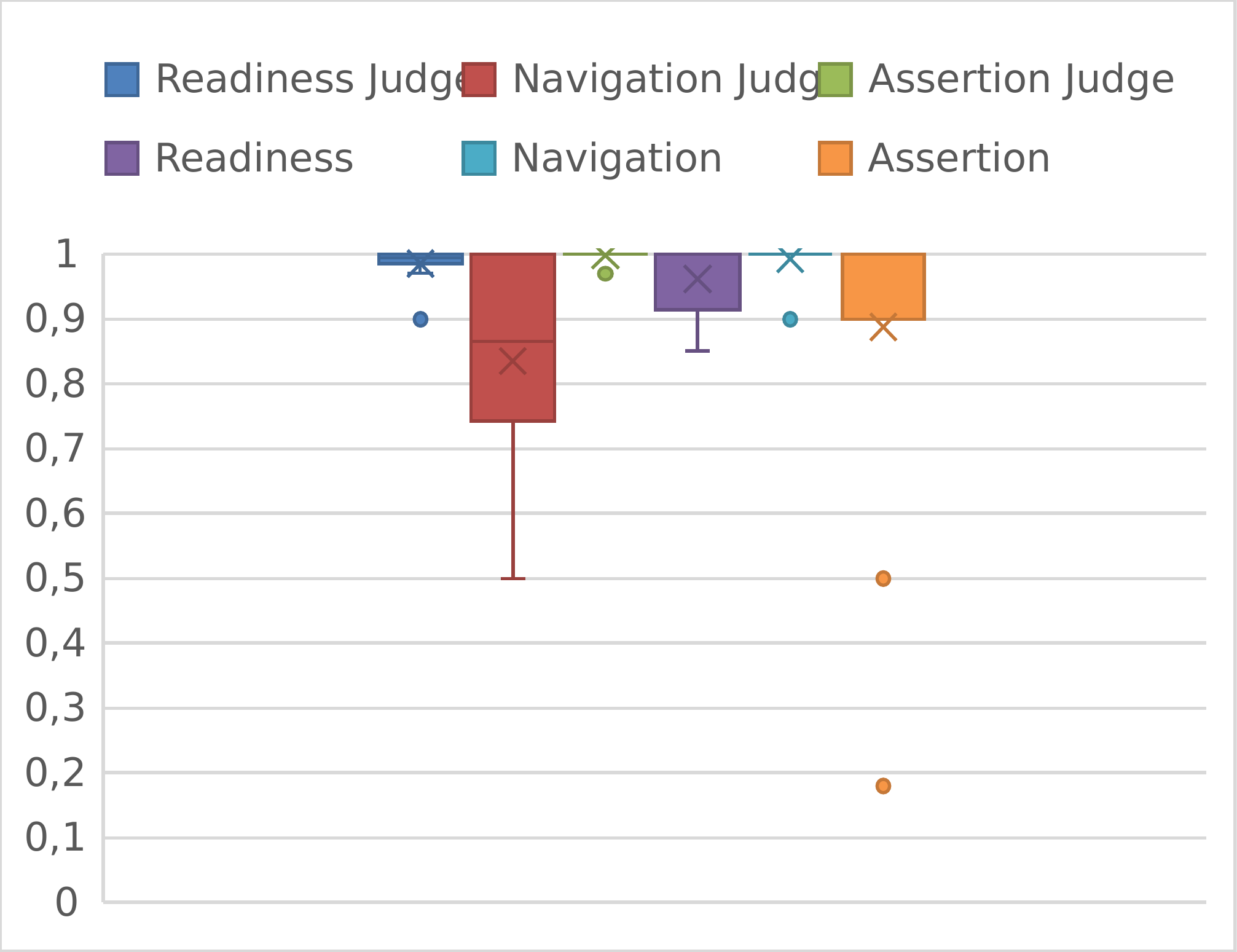}
		\caption{Qwen 2.5 7B}
		\label{fig:sub6}
	\end{subfigure}
\caption{Top figures: comparison of LLM performance on the 5 capabilities. 
Bottom figures: comparison of LLM agent capabilities with (left bars) and without (right bars) verified navigation. Mean accuracies 
are averaged over 20 runs. 
}
\label{fig:eval-comparison}
\end{figure}

For the first experiment, each LLM is evaluated independently on each capability through repeated executions of the corresponding agent. This allows us to estimate both its mean success rate and consistency, producing a capability profile for each LLM. Figure~\ref{fig:eval-comparison} shows substantial performance differences across LLMs and capabilities. In particular, performance varies not only between models but also between click and fill operations for the same model. For example, Mistral-nemo 12B, 
which is a small and fast model, achieves a 100\% success rate on readiness and fill navigation but is less effective on the other capabilities. 
This suggests assigning it to the former agents and selecting stronger LLMs, based on the measured profiles. 
Future work could investigate evaluating different coalitions of agents end to end. 

The second experiment quantifies how far capability measurements obtained without verification diverge from those obtained with verified navigation. 
We use the three-agent configuration and compare repeated evaluations with and without verified navigation. For the trusted agent, we use Llama3.3 70B, which is larger the the evaluated models and the highest performing LLM given in~\cite{DBLP:conf/enase/SalvaT26}.  Figure~\ref{fig:eval-comparison} shows that 
measurements without verification deviate from verified ones by up to $44\%$ (see 
navigation with Qwen2.5 3B). 
This indicates that silent navigation failures can significantly affect the 
evaluations and that context replacement provides a more stable basis for comparing LLM capabilities. Additional results are given in the appendix.

\bibliographystyle{apalike}
{\bibliography{aaai2027} }

\end{document}


\appendix

\section{Additional Results}

Figure~\ref{fig:capability-comparison} shows the mean accuracy and consistency over 20 runs of the 9 evaluated LLMs on the five capabilities. 
To illustrate how these profiles can guide configuration, we select for each capability the LLM with the highest mean accuracy, breaking ties by smaller model size. 
Table~\ref{tab:assignment} shows the
resulting assignment. 

\begin{table}[h]
\centering
\small
\begin{tabular}{lll}
\hline
Agent & Selected LLM & Mean acc. \\
\hline
$\mathit{agent}^{c}_{read}$ & Gemma 4 e4b & 1.00 \\
$\mathit{agent}^{f}_{read}$ & Gemma 4 e4b & 1.00 \\
$\mathit{agent}^{c}_{nav}$ & Qwen 3.6 27B & 0.999 \\
$\mathit{agent}^{f}_{nav}$ & Mistral-nemo 12B & 1.00 \\
$\mathit{agent}_{assert}$ & Qwen 3.6 27B & 0.97 \\
\hline
\end{tabular}
\caption{Assignment suggested by the measured capability profiles
(highest mean accuracy per capability. Ties broken by smaller model size).}
\label{tab:assignment}
\end{table}

\begin{figure*}[t]
	\centering
	\begin{subfigure}[b]{0.25\textwidth}
		\centering
		\includegraphics[width=\textwidth]{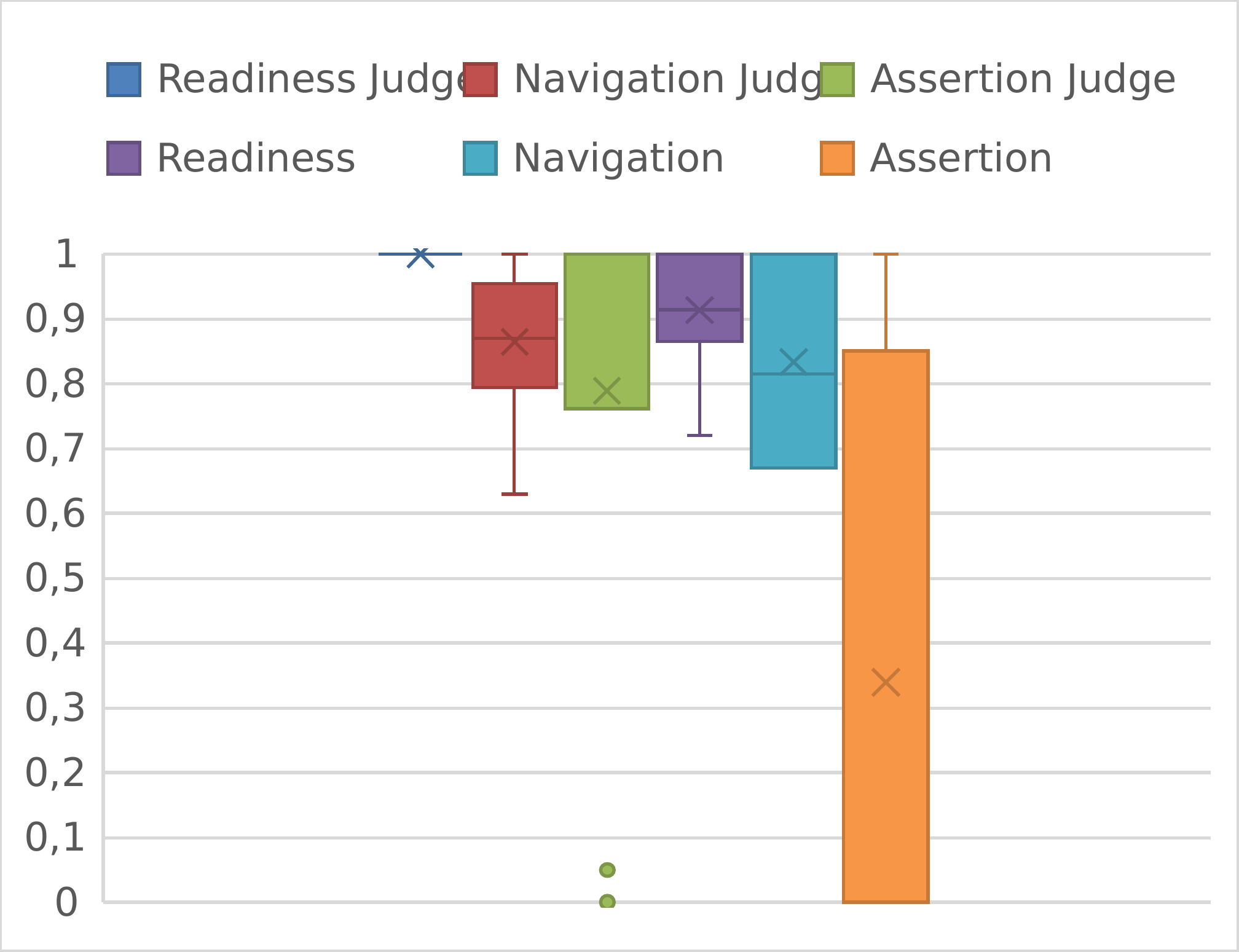}
		\caption{Mistral Nemo 12B}
		\label{fig:sub1}
	\end{subfigure}
	\hfill
	\begin{subfigure}[b]{0.25\textwidth}
		\centering
		\includegraphics[width=\textwidth]{Figures/Mistral-7b.pdf}
		\caption{Mistral 7B}
		\label{fig:sub2}
	\end{subfigure}
	\hfill
	\begin{subfigure}[b]{0.25\textwidth}
		\centering
		\includegraphics[width=\textwidth]{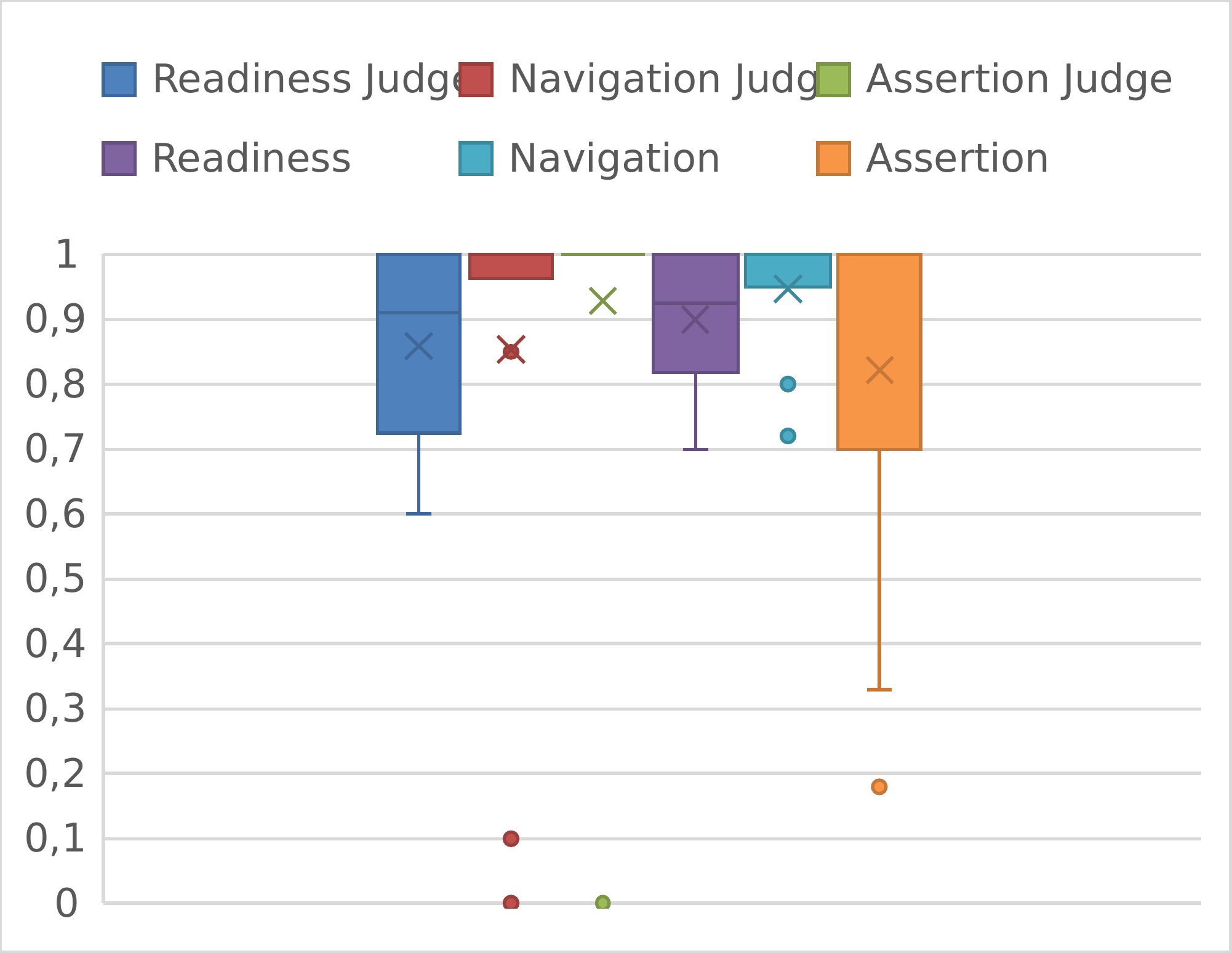}
		\caption{Mistral Devstral 24B}
		\label{fig:sub3}
	\end{subfigure}
	\hfill
	\begin{subfigure}[b]{0.25\textwidth}
		\centering
		\includegraphics[width=\textwidth]{Figures/qwen-2_5-3b.pdf}
		\caption{Qwen 2.5 3B}
		\label{fig:sub4}
	\end{subfigure}
	\hfill
	\begin{subfigure}[b]{0.25\textwidth}
		\centering
		\includegraphics[width=\textwidth]{Figures/qwen-2_5-7b.pdf}
		\caption{Qwen 2.5 7B}
		\label{fig:sub5}
	\end{subfigure}
	\hfill
	\begin{subfigure}[b]{0.25\textwidth}
		\centering
		\includegraphics[width=\textwidth]{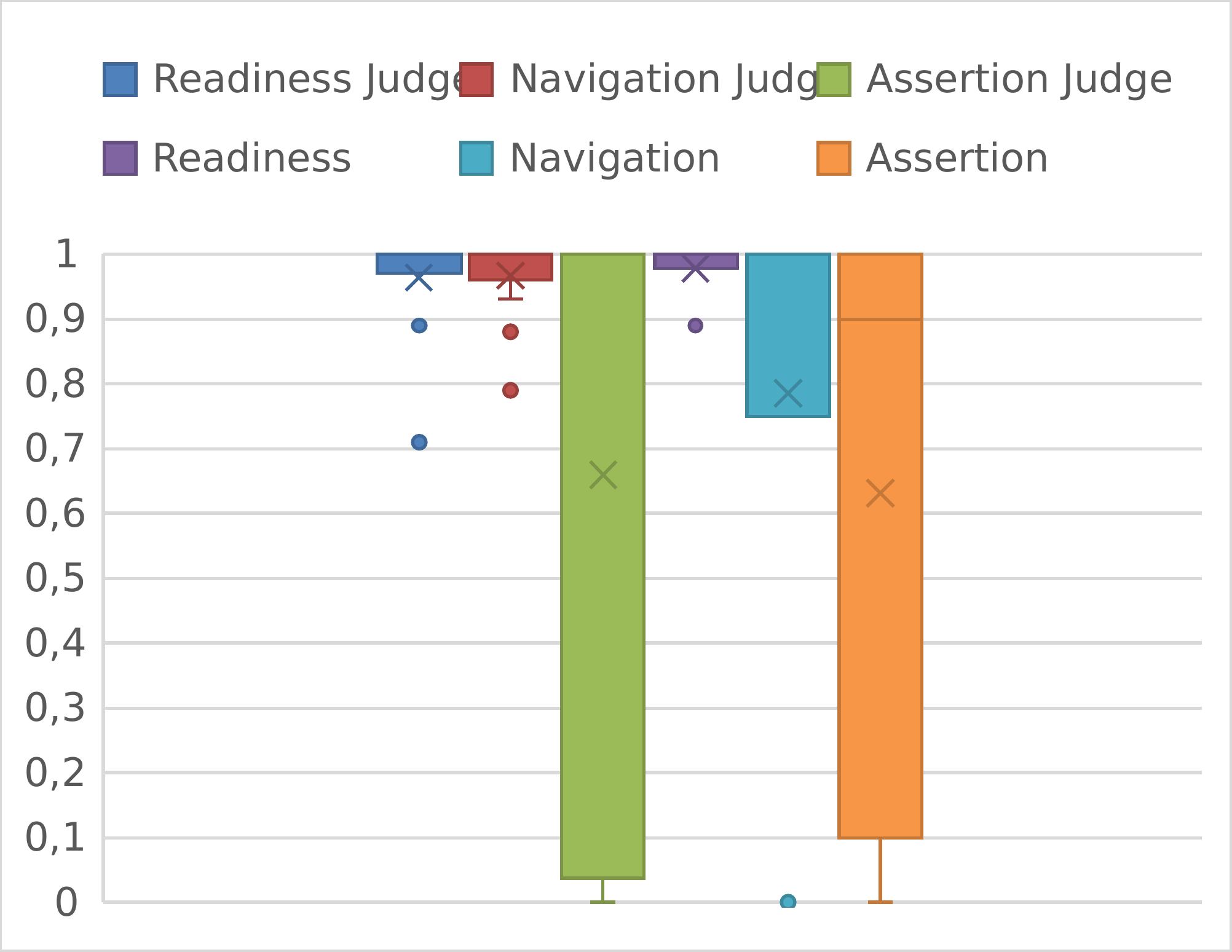}
		\caption{Qwen 3 14B}
		\label{fig:sub6}
	\end{subfigure}
\caption{Comparison of LLM agent capabilities with (left bars) and without (right bars) verified navigation. Mean accuracies 
are averaged over 20 runs. 
}
\label{fig:eval-comparison}
\end{figure*}

Figure~\ref{fig:eval-comparison} shows the difference between measurements obtained with and without verified navigation for 6 LLMs.

\begin{figure*}[h]
	\centering
	\begin{subfigure}[b]{0.3\textwidth}
		\centering
		\includegraphics[width=\textwidth]{Figures/5 agents devstral 24b.pdf}
		\caption{Devstral 24B}
		\label{fig:sub1}
	\end{subfigure}
	\hfill
	\begin{subfigure}[b]{0.3\textwidth}
		\centering
		\includegraphics[width=\textwidth]{Figures/5 agents Gemma4 e4b.pdf}
		\caption{Gemma 4 e4b}
		\label{fig:sub2}
	\end{subfigure}
	\hfill
	\begin{subfigure}[b]{0.3\textwidth}
		\centering
		\includegraphics[width=\textwidth]{Figures/5 agents  granite3,1 3b.pdf}
		\caption{Granite 3.1 3B}
		\label{fig:sub3}
	\end{subfigure}
	\hfill
	\begin{subfigure}[b]{0.3\textwidth}
		\centering
		\includegraphics[width=\textwidth]{Figures/5 agents  granite3,1 8b.pdf}
		\caption{Granite 3.1 8B}
		\label{fig:sub4}
	\end{subfigure}
	\hfill
	\begin{subfigure}[b]{0.3\textwidth}
		\centering
		\includegraphics[width=\textwidth]{Figures/5 agents  Mistral 7b.pdf}
		\caption{Mistral 7B}
		\label{fig:sub5}
	\end{subfigure}
	\hfill
	\begin{subfigure}[b]{0.3\textwidth}
		\centering
		\includegraphics[width=\textwidth]{Figures/5 agents  Mistral-nemo.pdf}
		\caption{Qwen 3 14B}
		\label{fig:sub6}
	\end{subfigure}
	\hfill
	\begin{subfigure}[b]{0.3\textwidth}
		\centering
		\includegraphics[width=\textwidth]{Figures/5 agents qwen3,6 27b.pdf}
		\caption{Qwen 3.6 27B}
		\label{fig:sub4}
	\end{subfigure}
	\hfill
	\begin{subfigure}[b]{0.3\textwidth}
		\centering
		\includegraphics[width=\textwidth]{Figures/5 agents gpt-oss 20b.pdf}
		\caption{Gpt-oss 20B}
		\label{fig:sub5}
	\end{subfigure}
	\hfill
	\begin{subfigure}[b]{0.3\textwidth}
		\centering
		\includegraphics[width=\textwidth]{Figures/5 agents Mistral-nemo.pdf}
		\caption{Mistral Nemo 12B}
		\label{fig:sub6}
	\end{subfigure}
\caption{Comparison of LLM agent performance on the 5 capabilities. 
Mean accuracies 
are averaged over 20 runs. 
}
\label{fig:capability-comparison}
\end{figure*}

\bibliography{aaai2027}